\documentclass[]{interact}

\usepackage[numbers,sort&compress]{natbib}
\bibpunct[, ]{[}{]}{,}{n}{,}{,}

\makeatletter
\def\NAT@def@citea{\def\@citea{\NAT@separator}}
\makeatother
\usepackage{url}

\usepackage[colorlinks=true, linkcolor=blue, citecolor=blue, urlcolor=blue]{hyperref}

\usepackage{graphicx}
\usepackage{booktabs}
\usepackage{multirow}
\usepackage{tabularx}
\usepackage{enumitem}
\usepackage{threeparttable}

\newcommand{\change}[1]{\textcolor{black}{#1}}

\begin{document}

\title{An Empirical Survey on the Early Adoption of DNS \change{Certification} Authority Authorization}
\author{
\name{Jukka Ruohonen\textsuperscript{a}}
\affil{\textsuperscript{a}Department of Future Technologies, University of Turku, Finland (\texttt{juanruo@utu.fi})}}
%

\maketitle

%
\begin{abstract}
A new \change{certification} authority authorization (CAA) resource record for the domain name system (DNS) was standardized in 2013. Motivated by the later 2017 decision to enforce mandatory CAA checking for most certificate authorities, this paper surveys the early adoption of CAA by using an empirical sample collected from the Alexa's top-million domains. According to the results, (i) the adoption of CAA is still at a modest level; only a little below two percent of the popular domains sampled have adopted CAA. Among the domains that have adopted CAA, (ii) authorizations dealing with wildcard certificates are rare compared to conventional certificates. Interestingly, (iii) the results only partially reflect the market structure of the global certificate business. With these timely results, the paper contributes to the ongoing large-scale empirical research on the use of encryption technologies.
\end{abstract}


\section{Introduction}




The CAA resource record (RR) passed formal standardization in 2013.\footnote{~This manuscript was submitted for double-blind peer review to \textit{Journal of Cyber Security Technology} (ISSN: 2374-2917) in November 6, 2017. After this submission, two relevant papers have been published: \cite{Amann17} and \cite{Scheitle18}. A reader should consult these papers in addition to this manuscript.} It was only four years later, in March 2017, when \change{many} certificate authorities (CAs) were mandated to implement CAA checking~\cite{CABROWSER17}. This decision was celebrated due to increasing concerns about the issuance of fraudulent or rogue certificates in the global public key infrastructure (PKI). Although CAA is only a small step toward better PKI security, these concerns are a good way to briefly motivate the paper's empirical survey on the adoption and use of the new CAA resource record among popular Internet domains.

The transport layer security (TLS) and its predecessor, the secure sockets layer (SSL), constitute the almost universally adopted cryptographic protocols for establishing communications security in the Internet and the world wide web in particular. These protocols have largely also survived the test of time. While many revisions and alterations have been required to patch the protocols, no cryptographic vulnerability is known to exist that would allow to fully compromise the underlying end-to-end security model. This said, a number of high-profile TLS/SSL vulnerabilities (including the so-called \change{Heartbleed}, FREAK, and DROWN) have been discovered and disclosed in recent years \cite{Oh16, RFC7457}. In addition to these vulnerabilities, a long-standing weakness originates from the global PKI and its CA-based trust model \cite{Forno01, Hunt01}. This type of a weakness is also what motivated the standardization and the later enforcement of CAA checking among most certificate authorities.

The TLS protocol uses public-key cryptography for the initial exchange of keys. If the exchange is successful, symmetric encryption is used afterwards. The PKI-induced weakness exposes itself through server authentication. When a client connects to a web server using TLS, the client's web browser verifies the server's certificate (key) by evaluating it against a set of trusted certificates signed by trusted but still third-party authorities. This evaluation places the trust provision entirely into the hands of certificate authorities. In principle, a CA can sign a certificate for any domain, and this certificate will be trusted by most web browsers and operating systems. There are hundreds of CAs located in almost every corner of the world, and all of these CAs are equally trusted \cite{Dacosta12}. What is more, the trust relationship is transitive: web browsers and operating systems will trust certificates issued also by intermediate authorities~\cite{Akhawe13}. These fundamental pillars of the global PKI allow to understand why it can be desirable to only trust some CAs and mistrust others. To put theoretical issues aside, there have been two tangential real-world weaknesses in this trust model.

On one hand, the history knows many cases of erroneous signing of certificates by CAs~\cite{Meyer14}. For instance, in 2001 two certificates were issued by VeriSign in Microsoft's name to an individual not affiliated with Microsoft~\cite{Forno01}. While VeriSign sold its certificate business to Symantec later in 2010, controversies continued. For instance, recently in March 2017, Google blamed Symantec for incorrectly issuing tens of thousands of certificates~\cite{BBC17}. A few months later, in August 2017, Symantec in turn sold its certificate business to DigiCert, but regardless of the ownership, the controversies are likely to continue about the rigor (or lack thereof) that CAs use for certificate issuance. 

On the other hand, the trust provision has exposed a more pressing concern about the security of the CAs themselves. Any CA is a lucrative target for attacks. In addition to limiting the availability of a CA through denial-of-service attacks~\cite{LinJing12}, compromising a CA opens numerous possibilities for tampering with confidentiality and integrity, including issuance of rogue certificates, stealing or substituting keys used for signing, and manipulating requests for issuing valid certificates for fraudulent purposes \cite{Grimm16, KimKwon17}. These scenarios are not only theoretical. The likely most well-known attack occurred in 2011 against the Dutch certificate authority DigiNotar, which later went also bankrupt due to the attack. By compromising the CA, an attacker was able to issue a wildcard certificate for Google's main domain, which was subsequently used to conduct man-in-the-middle attacks against users in Iran~\cite{Leavitt11}. The same pattern had already occurred a few months earlier when another certificate authority, Comodo, was compromised. Both cases led to speculations about the potential involvement of state-level actors. Such speculations have also intensified in recent years due to numerous legislative changes made to increase the power of national intelligence agencies \cite{Dacosta12}. For instance, concerns were raised in October 2017 about the implications of a Dutch law for trusting the national, state-level CA of the Netherlands~\cite{TheRegister17a}. On the bright side, the attacks and the controversies led to many initiatives for improving the security of the global PKI.

The current initiatives for improving TLS/PKI security are too voluminous~\cite{WeiWulf17} to adequately summarize in the present context. Nevertheless---insofar as the certificate issuance problems are considered, the standardization of the so-called certificate transparency~\cite{RFC6962} is noteworthy because it enables auditing of CAs through publicly available logs. Although there is still room for improvements, the transparency logs have also proven useful for better understanding the Internet-wide TLS/PKI ecosystem~\cite{VanderSloot16}. Given the increasing concerns about the ever increasing mass surveillance~\cite{RFC7258, Schuster17}, also the so-called Let's Encrypt (LE) initiative is worth remarking. By offering a free certificate and easy configuration options, the initiative is noteworthy as it attempts to democratize certificate issuance by challenging both the state-led mass surveillance endeavors and the global certificate business~\cite{Aertsen17}. The initiative also supports and respects \change{certification} authorization~\cite{LetsEncrypt17}, which is one of the standardized initiatives for improving TLS/PKI security. Before proceeding to discuss CAA in detail, it should be emphasized that the rationale behind CAA is to address the first weakness type, the erroneous issuance of certificates. This rationale is implemented through DNS by allowing any domain name owner to restrict the issuance of certificates by CAs.

\section{\change{Certification} Authority Authorization}

The CAA resource record type was created to allow a domain name owner to specify those certificate authorities who are allowed to issue certificates for the domain name. Before issuing certificates for domains, CAs evaluate the presence of CAA records via live DNS resolving. If the records are either missing or the records specifically authorize a given CA, the certificate authority is permitted to proceed with the certificate issuance. In essence, therefore, CAA provides an additional assertion to prevent erroneous or accidental issuance of certificates. This DNS-based authorization mechanism neither addresses nor prevents threats originating from the insecurity of certificate authorities. That is\change{:} by compromising a CA, it is presumably trivial to also bypass CAA checking when issuing rogue certificates or conducting some related attacks.

The RFC 6844 specification \cite{RFC6844} for CAA is fairly simple. The RR structure contains flags, tags, and values. Currently, only the so-called critical flag is defined. Specifying this flag instructs a CA to not issue a certificate in case the CA does not implement or understand the semantics of a given tag-value pair. There are three valid tags: (1) the \textit{issue} tag is used to define a certificate authority who is authorized to issue certificates for the domain in question; (2) the \textit{issuewild} tag is similar but authorizes the issuance of wildcard certificates; and (3) the \textit{iodef} tag enables the use of a standardized message exchange format~\cite{RFC5070} to report CAA-related problems to the owner of the domain. For the \textit{issue} and \textit{issuewild} tags, the value specifies the CA who is authorized. An empty value is interpreted to disallow the issuance. For instance, the following imaginary record disallows the issuance of wildcard certificates, while permitting DigiCert to issue normal certificates and instructing CAs to use electronic mail for reporting problems:
\begin{quote}
\begin{verbatim}
     example.com    CAA 0 issuewild ";"
     example.com    CAA 0 issue "digicert.com"
     example.com    CAA 0 iodef "mailto:root@example.com"
\end{verbatim}
\end{quote}

Despite of this simple format, many DNS servers and tools do not yet support the CAA resource record. In addition to this lack of supporting software, the current adoption challenges relate to domain name aliases, the lack of clearly defined accounting criteria on how CAs are evaluating CAA records in practice, and configuration problems for multinational companies who may control hundreds of domains and their subdomains \cite{Rowley17}. While many of these problems likely continue to hinder adoption, the question about domain name aliases is directly relevant also for this paper. In other words, a few details about the empirical sample should be elaborated before proceeding to evaluate the adoption of the \change{certification} authority authorization.

\section{Data}

The empirical sample is based on the Alexa's top-million busiest domains \cite{Alexa17}. Although some skepticism is warranted about the representativeness of the list particularly in the DNS context \cite{Wander17}, the list is commonly used as a benchmark in large-scale empirical studies exploring TLS/PKI~\cite{Springall16, VanderSloot16}, among other Internet measurement topics~\cite{Delamore15, Ruohonen17APSEC}. Because the paper's focus is restricted to DNS, a simple client-side DNS resolver~\cite{Ruohonen16COMPSYSTECH} \change{was} used to query for CAA resource records of each domain in the Alexa's popularity list using Google's name server at \texttt{8.8.8.8}. This querying requires a brief elaboration.

Each domain in the list was resolved three times in order to rule out timeouts and other temporary resolution failures. The same applies to further resolving required to process the CAA records from the viewpoint of certificate authorities~\cite{RFC6844}. For dealing with aliases (CNAMEs), a generally recommended upper limit~\cite{RFC1536} is used by fixing the maximum number of recursive queries to eight. This recursion depth is also noted in the errata for the CAA standard \cite{RFC6844Errata}. As an example: if the CAA records of a domain \texttt{a.b.c} are aliased to \texttt{d.e.f}, which in turn has a CNAME pointing to \texttt{g.h.i}, after two recursive queries (for resolving \texttt{d.e.f} and \texttt{g.h.i}), the CAA records of \texttt{g.h.i} are used for \texttt{a.b.c}, provided that these are present. If this recursive search does not find CAA records, the records are queried by moving upward in the DNS hierarchy until the top-level domain (TLD) name is reached~\cite{RFC6844}. For instance: if no CAA records are found by traversing the CNAME chain of \texttt{a.b.c}, the CAA records of \texttt{b.c}, if present, are used for the domain \texttt{a.b.c}. If also this hierarchical query fails, the domain \texttt{a.b.c} is finally concluded to operate without CAA records, given that the subsequent hierarchical level denotes a top-level domain to which the hierarchical querying does not traverse.

A brief semantic validation is carried out for each CAA record. Namely: (1) each resource is verified to include a flag, a tag, and a value; (2) flags are asserted to equal $128$, \change{one, or} zero; (3) tags are validated to be either \textit{issue}, \textit{issuewild}, or \textit{iodef}; and (4) values specified with \textit{iodef} are checked to start either with the character string \texttt{mailto} or with the string \texttt{http}. These basic checks caught a few small errors in the specifications of the CAA records among the domains sampled. For instance, a few domains omit the \texttt{mailto}, \texttt{http}, or \texttt{https} prefixes when specifying values for the \textit{iodef} tag. Also other innocent mistakes are present: the domains \texttt{manned.org}, \texttt{thousandwonders.net}, \texttt{vncg.org}, and \texttt{yorhel.nl} have tried to specify wildcard authorizations but spelled the corresponding tag as \texttt{issuewold}. Likewise, the domain \texttt{globo.com} has spelled the \textit{iodef} tag as \texttt{ideof}, and the domain \texttt{hivolda.no} has specified two flags. While these few cases are statistical outliers, the cases are excluded to ensure consistent parsing. It is worth to also remark that it would be possible to carry out further validation by resolving also the authorized domains, including the mail exchange (MX) records for the email addresses delivered via the \textit{iodef} tag. As the focus of the paper is on the adoption of CAA, such validation experiments can be left for further work, however.

\section{Results}

\begin{figure}[th!b]
\centering
\includegraphics[width=12cm, height=3.5cm]{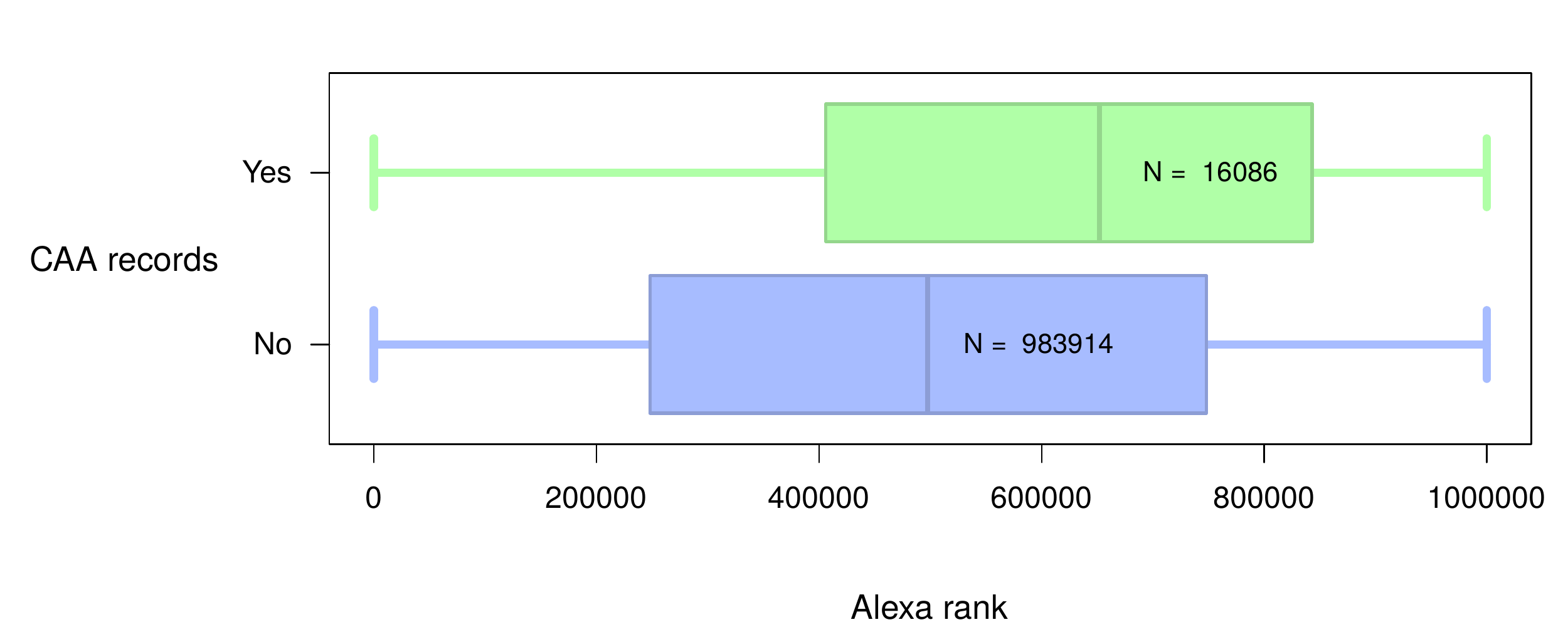}
\caption{Presence of CAA Records and Alexa's Popularity Ranks}
\label{fig: alexa}
\end{figure}

\begin{figure}[th!b]
\centering
\includegraphics[width=12cm, height=4cm]{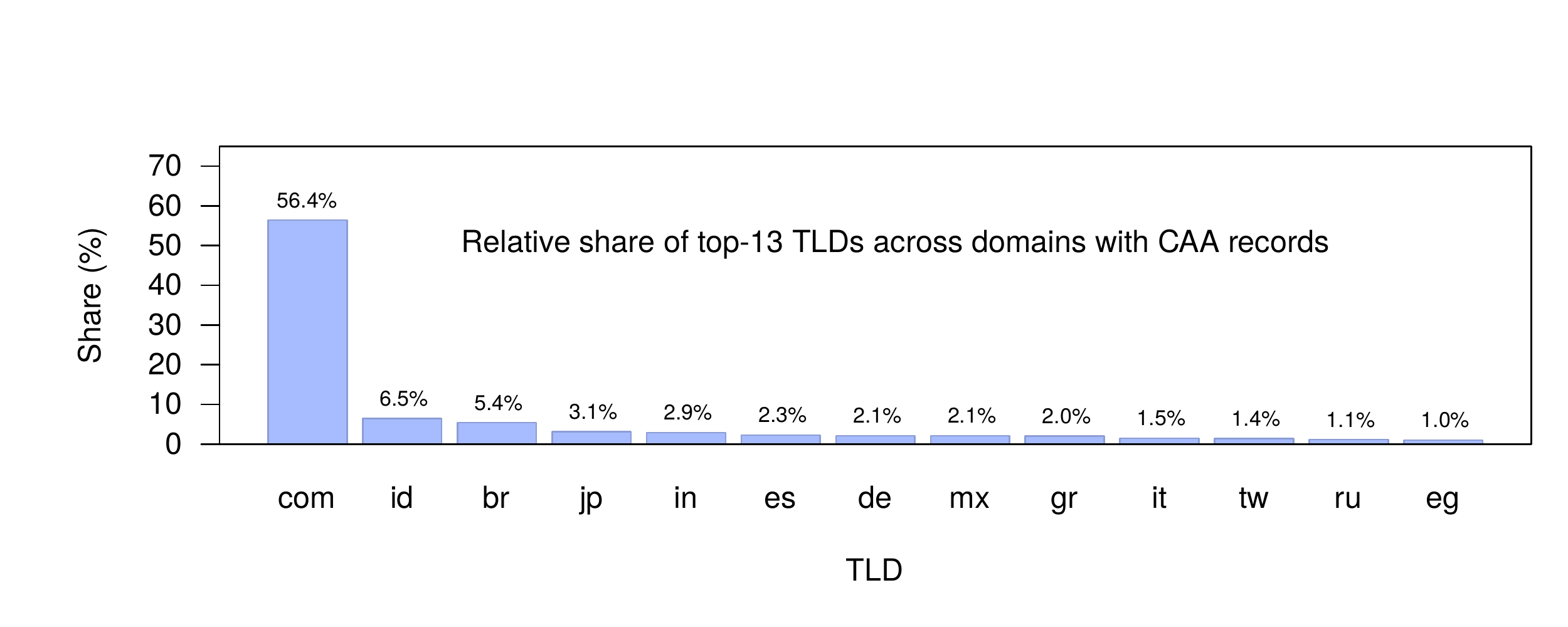}
\caption{Domains with CAA Records According to TLDs}
\label{fig: tlds}
\end{figure}

The dataset obtained via live DNS resolving contains $16086$ domains with CAA resource records. Thus, in absolute terms, the adoption is still at a rather modest level: only about 1.6\% of the domains in the Alexa's top-million list have specified \change{certification} authority authorizations in their DNS records. Even though existing scholarly research is limited, there are good reasons to suspect that the pace of adoption has increased after the CAA checking was voted as mandatory for certificate authorities participating in the CA/Browser Forum. For instance, only 307 domains out of Alexa's top-million list were reported to have deployed CAA records in early April~2017 \cite{Aleksandersen17}. If this amount is used as a coarse heuristic, the adoption has increased by over 5000\% in about six months. To a small extent, the adoption has been slower among more popular domains compared to less popular domains. This observation is illustrated in Fig.~\ref{fig: alexa}, which shows the presence of CAA records across Alexa's popularity rank. Although \texttt{com} is the most frequent TLD among the sampled domains with CAA records, also many country-code TLDs appear in the ranking shown in Fig.~\ref{fig: tlds}.

\begin{figure}[th!b]
\centering
\includegraphics[width=12cm, height=3.5cm]{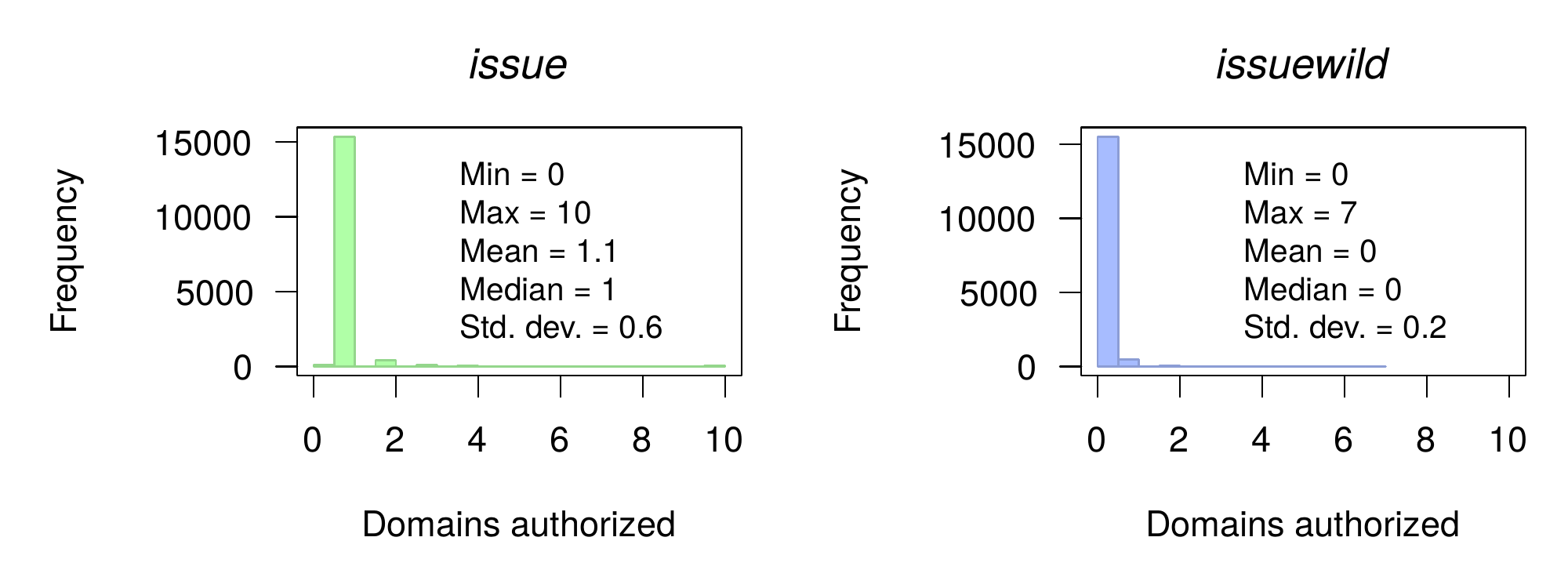}
\caption{Domains Authorized}
\label{fig: authorizations}
\end{figure}

On average, most of the domains with CAA records authorize only one CA. Most of these authorizations authorize CAs to issue only normal certificates. Authorizations involving wildcard certificates are less common (see Fig.~\ref{fig: authorizations}). The so-called DNS graphs offer a good method to elaborate these observations further. 

\begin{figure}[p!]
\centering
\includegraphics[width=12cm, height=7cm]{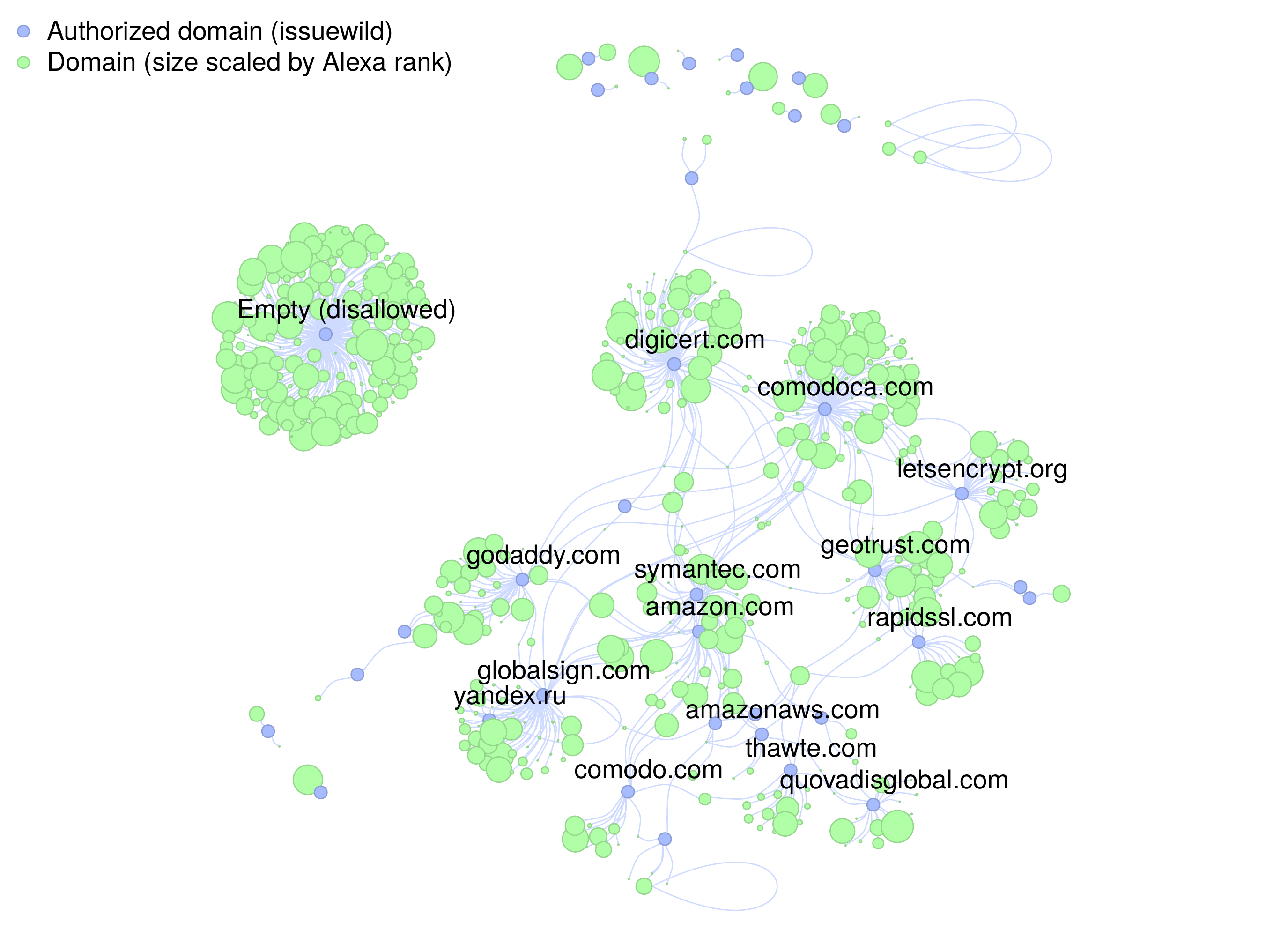}
\caption{Bipartite DNS-CAA Graph According to Wildcard Authorization}
\label{fig: issuewild}
%
\vspace{20pt}
%
\centering
\includegraphics[width=12cm, height=8cm]{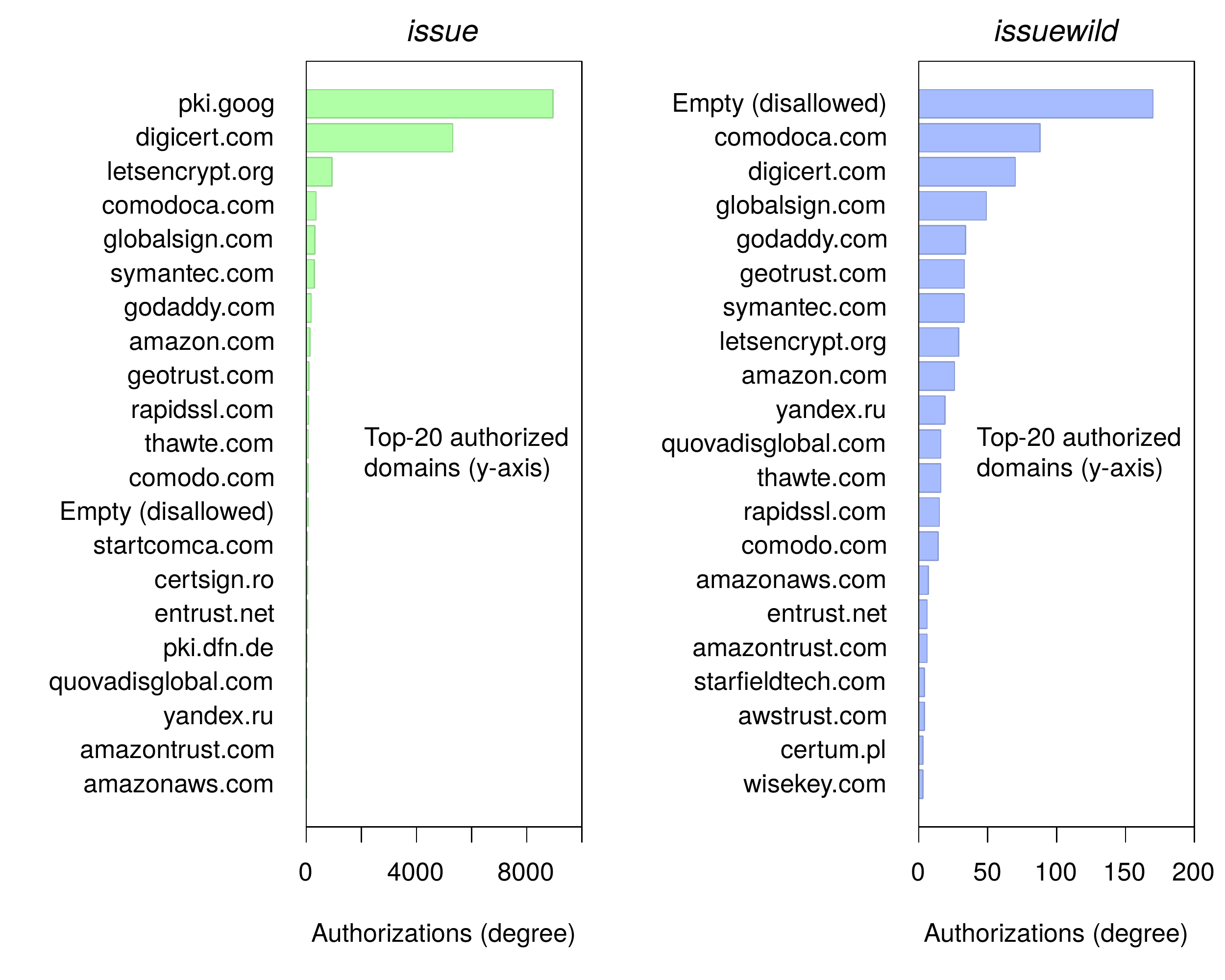}
\caption{Most Frequently Authorized Domains}
\label{fig: degrees}
%
\vspace{20pt}
%
\centering
\includegraphics[width=12cm, height=3cm]{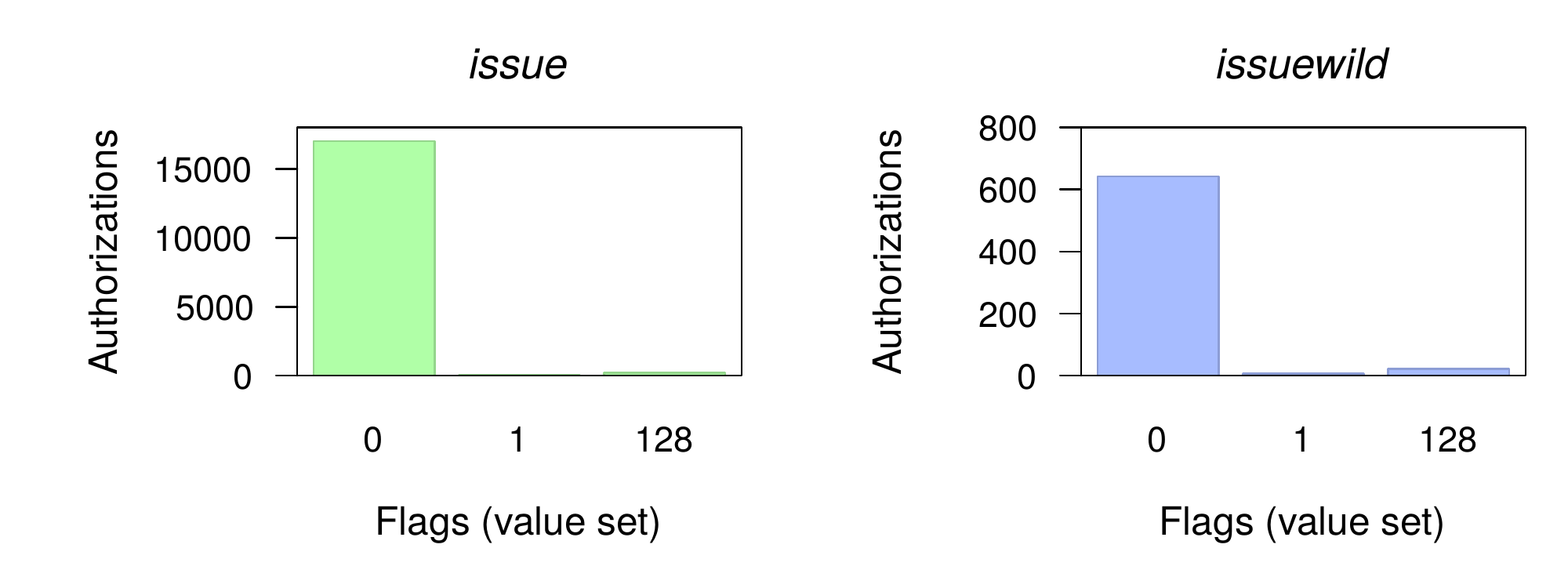}
\caption{Flags Set for the Authorizations}
\label{fig: flags}
\end{figure}

By using a common undirected and bipartite DNS graph representation~\cite{Ruohonen17CIT}, there are two types of vertices: the domains sampled from the Alexa's list and the domains the sampled domains have authorized via the \textit{issue} or the \textit{issuewild} tags. Whenever a given ``Alexa-domain'' has authorized a ``CA-domain'', an edge is placed between these two types of domain vertices. A resulting graph representation is illustrated in Fig.~\ref{fig: issuewild} by connecting the domains sampled (as represented by the green vertices with their sizes proportional to Alexa's popularity ranks) to the CA-domains (as represented by blue vertices) that the sampled Alexa-domains have authorized to issue wildcard certificates. By constructing an analogous bipartite graph according to the \textit{issue} tag, the most frequently authorized domains can be summarized by calculating the number of adjacent vertices to the Alexa-domains. The results from this calculation are shown in Fig.~\ref{fig: degrees}. There are many relevant observations to make from these two figures.

\begin{figure}[p!]
\centering
\includegraphics[width=13cm, height=13cm]{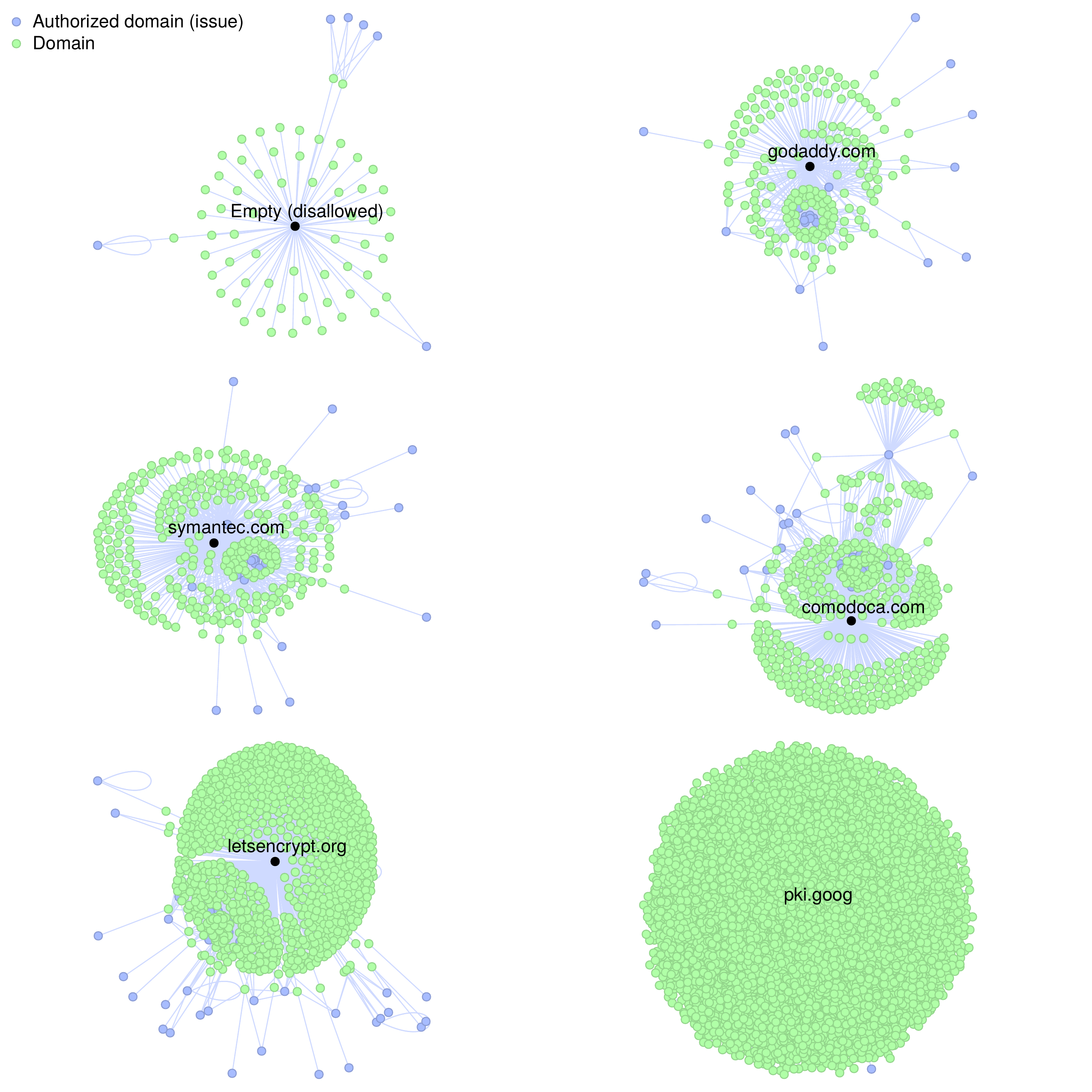}
\caption{Six DNS-CAA Subgraphs (two hops from the labels shown, \textit{issue})}
\label{fig: subgraphs}
%
\vspace{20pt}
%
\centering
\includegraphics[width=12cm, height=7cm]{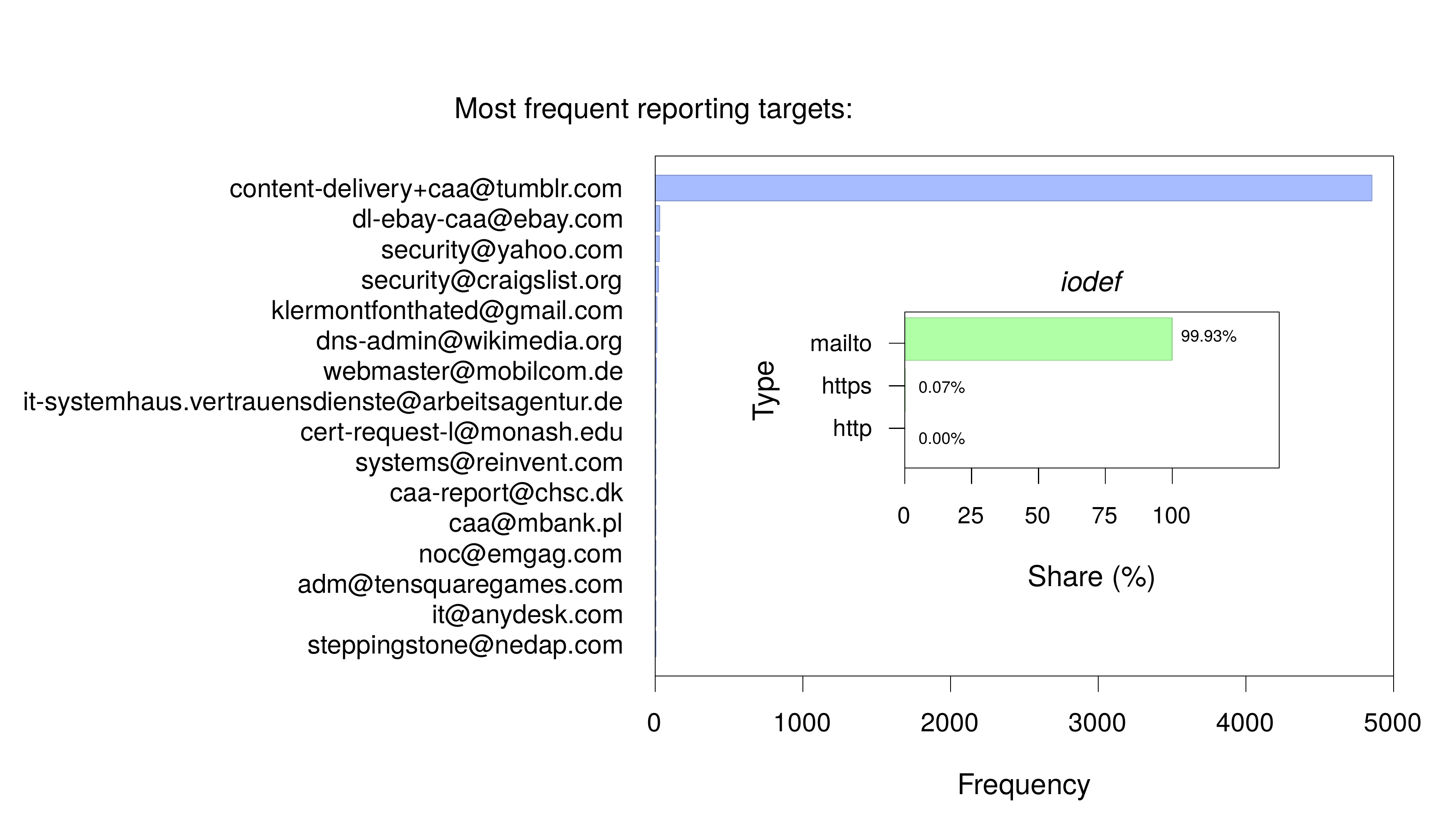}
\caption{A Summary of Reporting Targets}
\label{fig: iodef}
\end{figure}

All of the big commercial CAs are represented, but the CAA authorizations do not entirely reflect the market \change{shares} of the CA companies~\cite{Netcraft15}. Interestingly, neither are the numerous acquisitions and other changes in the market structure (yet) reflected in the results. What is more important, the most frequently authorized domain in the \textit{issue} graph is Google's \texttt{pki.goog}. This observation does not mean that many domains would authorize Google to issue certificates. The explanation rather is that many of Google's domains are popular, and, hence, present also in the Alexa's top-million list. A closer look reveals that out of the $8954$ domains connected to \texttt{pki.goog} in the \textit{issue} graph, as much as $8574$ contain the character string \texttt{blogspot}. These cases also reflect the resolving strategy; the CAA records for most of these cases were resolved via DNS hierarchy. In fact, out of the $16086$ domains with CAA records, only three were obtained via the CNAME traversing, about 14\% were directly available, and as much as 86\% were retrieved by moving upward in the DNS hierarchy. A majority of the hierarchical resolving cases refer to Google's domains.

The large disconnected subgraph in Fig.~\ref{fig: issuewild} represents domains that have disallowed the issuance of wildcard certificates for all authorities by specifying empty values for their \textit{issuewild} tags. Given the concerns about erroneous certificate issuance, this observation can be seen as a positive finding. However, all of these $170$ domains have accompanied their empty \textit{issuewild} tags with non-empty \textit{issue} tags, meaning that the domains have still authorized CAs to issue normal certificates. It is also noteworthy that the amount of self-loops is only eight and fifteen in the \textit{issue} and \textit{issuewild} graphs, respectively. In other words, only a few \change{domains} have authorized themselves. While \texttt{digicert.com} is among these domains, most of these cases likely refer to deployments operating with self-signed certificates. 

It is interesting to observe that many of the authorized CAs tend to cluster into their own dense subgraphs that are only sparsely connected to other clustered subgraphs. While this observation is a typical finding in empirical graph mining applications~\cite{Ruohonen16AICCSA}, the contextual interpretation is more interesting than statistical clustering as such. To aid the interpretation, Fig.~\ref{fig: subgraphs} displays six subgraphs that include those vertices and edges that are two steps away from the labeled domains authorized via the \textit{issue} tag. 

The presence of dense subgraphs means that most Alexa-domains tend to only authorize one CA-domain, which is understandable because most certificates cost money. Interestingly, however, the domains that have authorized \texttt{letsencrypt.org} via the \textit{issue} tag tend to form a more dense subgraph than many of the domains that have authorized commercial CAs. Even though LE is less frequently authorized than DigiCert \change{(see Fig.~\ref{fig: degrees})}, for instance, this result likely again reflects the sampling from the Alexa's list. In other words, the most popular domains are not the primary target audience of the LE initiative. The denseness of the LE-subgraph in Fig.~\ref{fig: subgraphs} likely reflects this target audience. In absolute terms, the results also implicitly correlate rather well with previous results according to which about 5\% of domains in the Alexa's top-million list used a LE's certificate in late 2016~\cite{Aertsen17}. While also the earlier remark about Google's domains is vividly reinforced by the subgraph illustration in Fig.~\ref{fig: subgraphs}, the upper-left subgraph is also noteworthy because it indicates the presence of a few configuration mistakes. That is, specifying both an empty value and a specified issuer is identical to specifying just the specified issuer alone \cite{RFC6844}. Analogous point can be made regarding the flags specified for authorizations; a few domains have set a value one rather than the value $128$ for specifying the critical flag (see Fig.~\ref{fig: flags}). Finally, most of the domains that have used the \textit{iodef} tag prefer electronic mail for reporting CAA-related issues (see Fig.~\ref{fig: iodef}). The majority of the domains with CAA records (about 65\%, to be precise) have not bothered to announce any contact details, however.

\section{Conclusion}


This paper surveyed the adoption of the \change{certification} authority authorization. By using a dataset based on live DNS resolving of domains in the Alexa's top-million list, the adoption of CAA was observed to still be at a modest level. Among the domains that have specified CAA records in their DNS configurations, the authorization of normal certificates is more common than the authorization of wildcard certificates. In fact, many domains explicit disallow CAs to issue wildcard certificates. In terms of the authorized CAs, the results reported reflect only partially the market structure of the global certificate business. In addition to new initiatives such as Let's Encrypt, particularly the domains owned by Google have pushed the adoption of CAA forward. These results align well with the continuing concerns about weaknesses in the global TLS/PKI ecosystem. \change{Four} limitations can be also noted for further work on CAA.

The paper's scope was strictly restricted to DNS. Therefore, (1) further research is required for better understanding how the DNS-based certificate authorizations correlate with the actual issuance of certificates by certificate authorities. The certificate transparency logs provide a good empirical source to examine such correlations. The topic is important because CAA does not impose any technique for validating whether a CA is properly evaluating CAA resource records~\cite{Rowley17}. Likewise, the use of DNS security (DNSSEC) is strongly recommended to be used in conjunction with CAA~\cite{RFC6844}. Although this conjunction was not examined in this paper, (2) it should be addressed in further empirical research. Nor did the paper attempt to track the adoption across time. Given that CAA checking was recently voted to be mandatory for most certificate authorities, (3) further research is also needed to systematically track CAA's future longitudinal evolution in different areas. For instance, an interesting topical area relates to the future adoption of CAA among content delivery networks and their DNS infrastructures. Finally, the sample used in the paper was restricted to the most popular domains. As this restriction presumably entails some biases for observing new TLS/PKI openings such as LE, (4) further work is required to evaluate different sampling strategies for large-scale empirical Internet measurement experiments.

\bibliographystyle{apalike}

\end{document}